\documentclass[prd,eqsecnum,twocolumn,amsfonts,amssymb]{revtex4}

\usepackage{graphicx}

\usepackage{bm}

\newcommand{\beq}{\begin{equation}}
\newcommand{\eeq}{\end{equation}}
\newcommand{\beqs}{\begin{eqnarray}}
\newcommand{\eeqs}{\end{eqnarray}}

\newcommand{\drawsquare}[2]{\hbox{%
\rule{#2pt}{#1pt}\hskip-#2pt
\rule{#1pt}{#2pt}\hskip-#1pt
\rule[#1pt]{#1pt}{#2pt}}\rule[#1pt]{#2pt}{#2pt}\hskip-#2pt
\rule{#2pt}{#1pt}}
\newcommand{\fund}{\raisebox{-.5pt}{\drawsquare{6.5}{0.4}}}
\newcommand{\sym}{\raisebox{-.5pt}{\drawsquare{6.5}{0.4}}\hskip-0.4pt%
        \raisebox{-.5pt}{\drawsquare{6.5}{0.4}}}
\newcommand{\asym}{\raisebox{-3.5pt}{\drawsquare{6.5}{0.4}}\hskip-6.9pt%
        \raisebox{3pt}{\drawsquare{6.5}{0.4}}}

\begin{document}

\title{Infrared Evolution and Phase Structure of a Gauge Theory
 Containing Different Fermion Representations}

\author{Thomas A. Ryttov}

\author{Robert Shrock}

\affiliation{
C. N. Yang Institute for Theoretical Physics \\
State University of New York \\
Stony Brook, NY 11794}

\begin{abstract}

We study the evolution of an asymptotically free vectorial SU($N$) gauge theory
from the ultraviolet to the infrared and the resultant phase structure in the
general case in which the theory contains fermions transforming according to
several different representations of the gauge group.  We discuss the
sequential fermion condensation and dynamical mass generation that occur, and
comment on the effect of bare fermion mass terms.

\end{abstract}


\maketitle

\section{Introduction} 

The phase structure of a non-Abelian gauge theory depends on its fermion
content.  Here we consider an asymptotically free vectorial gauge theory (in
$(3+1)$ dimensions, at zero temperature and chemical potential) with an SU($N$)
gauge group and fermions corresponding to several different representations of
the gauge group.  We denote the running gauge coupling of the theory as
$g(\mu)$, with $\alpha(\mu)=g(\mu)^2/(4\pi)$, where $\mu$ is the Euclidean
energy/momentum scale (which will often be suppressed in the notation).  Since
the gauge interaction is asymptotically free, at a sufficiently high energy
scale $\mu$, $\alpha(\mu)$ is small and the theory is perturbatively
calculable.  We will study a theory which contains several Dirac fermions
transforming according to different representations of SU($N$). We denote a
representation as $R$, the set of fermion representations in the theory as
$\{R\} \equiv \{R_1,...,R_k\}$, the number of Dirac fermions in each
representation $R_i$ as $N_{R_i}$, and the set of these numbers as $\{N_R\}
\equiv \{N_{R_1},...,N_{R_k}\}$ \cite{rn}.  We will first consider the case in
which all of these fermions are massless or have bare masses in the high-scale
Lagrangian that are small compared with the scale where $\alpha$ grows to a
size of order unity and the theory becomes strongly coupled.  One interesting
property of this type of theory is that it can exhibit fermion condensations at
different energy scales, with fermions with larger quadratic Casimir invariants
condensing and gaining dynamical masses at higher scales.  This theory could
arise from a larger one which is a chiral gauge theory, in which fermion masses
would generically be forbidden.  However, if we consider the theory by itself,
then, since it is vectorial, and hence fermion mass terms do not violate the
SU($N$) gauge symmetry, it is natural to consider a more complicated situation
in which some fermions have masses that are comparable to or greater than the
scale where the coupling $\alpha$ grows to O(1).  We shall also briefly comment
on this latter possibility.

   Although our work is an abstract field-theoretic study, not an effort to
construct a phenomenological model, we note that there has been considerable
interest recently in the analysis of vectorial non-Abelian gauge theories with
fermions in higher-dimensional representations, partly motivated by technicolor
model-building \cite{higherrep}-\cite{sanrev}.  We note in passing that in the
early development of the Standard Model, the possibility was considered that
the color SU(3)$_c$ sector might contain not just quarks but also other
fermions transforming as higher-dimensional representations of the color group
\cite{colorrep}.  Fermions in higher-dimensional representations have also been
used in constructions of chiral gauge theories, but here we restrict our
consideration to vectorial gauge theories.

\section{General Theoretical Framework}

\subsection{Beta Function} 

In this section we review the general theoretical framework that we will use in
our calculations.  The beta function of the theory is denoted $\beta = dg/dt$,
where $dt = d\ln \mu$. In terms of $\alpha$, this can be written as 
\beq
\frac{d\alpha}{dt} = - \frac{\alpha^2}{2\pi} \left [ b_1 + \frac{b_2 \, 
    \alpha}{4\pi} + O(\alpha^2) \right ]
\label{beta}
\eeq
where the coefficient $b_\ell$ arises at $\ell$-loop order in perturbation
theory, and the first two coefficients, $b_1$ and $b_2$, are
scheme-independent.  These are \cite{b1}
\beq
b_1 = \frac{1}{3}\left [ 11 C_2(G) - 4\sum_R N_R \, T(R) \right ]
\label{b1}
\eeq
and \cite{b2} 
\beq
b_2=\frac{1}{3}\left [ 34 C_2(G)^2 - 4\sum_R (5C_2(G)+3C_2(R))N_R \, 
T(R)\right ]
\ . 
\label{b2}
\eeq
Here $C_2(R)$ is the quadratic Casimir invariant and 
$T(R)$ is the trace invariant for the representation $R$ \cite{gp}, with 
$C_2(G) \equiv C_2(adj.)$ and $C_2({\rm SU}(N))=N$ (see appendix).  
The condition that the theory be asymptotically free, i.e., that $b_1 > 0$, 
yields the upper bound 
\beq
\sum_R N_R \, T(R) < \frac{11N}{4} \ . 
\label{afcondition}
\eeq
Since all of the terms on the left-hand side contribute positively, this
implies the upper bound on the number of fermions in each representation 
$N_R < N_{R,max}$, where 
\beq
N_{R,max} = \frac{11N}{4T(R)} \ . 
\label{nrmax} 
\eeq
Here and below, we implicitly carry out an analytic continuation of $N_R$ from
non-negative integers to non-negative real numbers; however, it is understood
that physically they are, of course, non-negative integers. If there are few
fermions, then also $b_2> 0$, so that the two-loop beta function has a zero
only at the origin, $\alpha=0$.  

 A sufficient increase in the numbers of fermions in various 
representations leads to a reversal in the sign of $b_2$ from positive to
negative, while still satisfying the condition of asymptotic freedom,
(\ref{afcondition}). For a set of fermion representations $\{N_R\}$ with this
property, the two-loop beta function has a zero away from the origin at
\beq
\alpha_{IR} = -\frac{4\pi b_1}{b_2} = \frac{4\pi b_1}{|b_2|}  \ . 
\label{alfir}
\eeq
For the theory with a single type of fermion representation, we denote the
value of $N_R$ where $b_2=0$ as $N_{R,IR}$.  This is
\beq
N_{R,IR} = \frac{17 C_2(G)^2}{2[5C_2(G)+3C_2(R)]T(R)} \ . 
\label{nrir}
\eeq
The fact that $N_{R,IR} < N_{R,max}$ is evident because for $N=N_{R,IR}$, $b_1$
has the positive (i.e., asymptotically free) value 
\beqs
b_1 & = & \frac{C_2(G) \, [6C_2(G)+11C_2(R) ]}{5C_2(G)+3C_2(R)} > 0 \cr\cr 
    & & {\rm for} \ \ N_R=N_{R,IR} \ . 
\label{b1pos_whereb2eq0}
\eeqs

If $b_2 < 0$, so that there is an infrared zero of the beta function, then as
the scale $\mu$ decreases from large values, $\alpha(\mu)$ increases toward
this value.  The infrared behavior then depends on whether or not the value of
the coupling $\alpha_{IR}$ is sufficiently large as to cause spontaneous chiral
symmetry breaking \cite{bz}. If the properties of the theory are such that no
fermion condensates form, then this is an exact infrared fixed point (IRFP) of
the (perturbatively calculated) renormalization group equation for $\alpha$.
If, on the other hand, some fermions do condense, so that they get dynamically
generated masses and are integrated out of the low-energy effective theory
applicable below the scale(s) of condensation, then, since the beta function
changes, the original value of $\alpha_{IR}$ is only an approximate IFRP.
Since the coefficients $b_1$ and $b_2$ are the maximal set of coefficients in
the beta function that are scheme-independent, it follows that conclusions
obtained from the two-loop beta function should be at least qualitatively
reliable physically.  However, since we will deal with values of $\alpha_{IR}$
of order unity, i.e., strongly coupled gauge interactions, it is understood
that there are inevitably significant theoretical uncertainties in the results.
In this context, we recall that the two-loop perturbative beta function is an
asymptotic expansion in $\alpha$ and does not include a number of important
effects, including confinement and instantons. Indeed, instanton effects
involve factors like $\exp(-c \pi/\alpha)$ (where $c$ is a constant), which
cannot be seen to any order of perturbation theory.  Moreover, it should be
noted that even if there is no zero of the two-loop beta function away from the
origin, i.e., a perturbative IRFP, the beta function may exhibit a
nonperturbative slowing of the running associated with the fact that at energy
scales below the confinement scale, the physics is not accurately described in
terms of the Lagrangian degrees of freedom (fermions and gluons)
\cite{lmax}-\cite{creutz}.  We observe that one can calculate $\alpha_{IR}$
more accurately using the higher-order coefficients of the beta function.
Finally, although an asymptotically free vectorial SU($N$) gauge theory of the
type that we consider here does not require an ultraviolet completion, it
could, as remarked above, arise as the low-energy effective field theory
resulting from the breaking of a larger, chiral, gauge symmetry.  In this case,
one would also want to assess the effects of residual higher-dimensional
operators from this larger gauge theory (e.g., \cite{4f}).

\subsection{Results from Approximate Solution of Dyson-Schwinger Equation for
  Fermion Propagator}

A solution of the Dyson-Schwinger (DS) equation for the propagator of a fermion
$\psi$ in the representation $R$ of the gauge group, with zero bare mass, in
the approximation of one-gluon (also called ladder) exchange, yields a nonzero,
dynamically generated mass if the coupling $\alpha(\mu)$ exceeds a critical
value $\alpha_{R,cr}$ given by \cite{lane}-\cite{bds}
\beq
\frac{3 C_2(R) \alpha_{R,cr}}{\pi} = 1 \ . 
\label{alfcrit}
\eeq
In the same ladder approximation, the anomalous dimension for the fermion
(bilinear) mass operator is $\gamma = 1$ at $\alpha = \alpha_{cr,R}$. Some
lattice studies have reported initial results on measurements of $\gamma$
\cite{afn,lgtsym}.  Corrections to the one-gluon exchange approximation have
been analyzed and found not to be too large \cite{alm}. To assess the
implications of these corrections for the boundary of the chirally symmetric
phase, one also calculates $\alpha_{IR}$ to the corresponding higher order.
Since the dynamically generated mass for this fermion is the coefficient of the
bilinear fermion operator in an effective Lagrangian, this indicates the
formation of a condensate of the fermions in the representation $R$, and
associated spontaneous chiral symmetry breaking (S$\chi$SB) by the gauge
interaction, as $\alpha$ increases through the critical value $\alpha_{R,cr}$
Some early studies with lattice simulations of chiral symmetry breaking were
carried out for SU(2) and SU(3) for various fermion representations in
\cite{ksssu2}-\cite{iwasaki}.  There has been considerable recent lattice work,
mainly on the group SU(3) with fermions in the fundamental representation or
rank-2 symmetric (sextet) representation and on SU(2) with fermions in the
adjoint (equivalent to rank-2 symmetric) representation.  Some of the rapidly
increasing number of papers reporting results from numerical lattice
simulations include Refs. \cite{afn}-\cite{lgtsym}. To our knowledge, there
have not been lattice studies of chiral symmetry breaking in a theory
containing dynamical fermions in two or more different representations
(simultaneously present).

The analysis of the gauge coupling evolution and chiral symmetry realization in
vectorial asymptotically free gauge theories has been of particular interest in
the context of technicolor (TC) theories \cite{tc}, especially in the context
of the most promising such theories, which exhibit a slowly running
(``walking'') gauge coupling associated with an approximate infrared fixed
point of the renormalization group \cite{wtc} (see also \cite{otherw}).  In the
actual application to theories of dynamical electroweak symmetry breaking, one
must embed the technicolor sector in a larger theory, extended technicolor
(ETC) in order to give masses to quarks and leptons and to account for their
generational structure \cite{etc}.  A necessary property of TC/ETC theories is
that the ETC symmetry must break in a series of stages to the TC symmetry,
which is an asymptotically free, vectorial theory that becomes strongly coupled
on the TeV scale, producing bilinear technifermion condensates that break the
electroweak gauge symmetry.  ETC is constructed as an asymptotically free
chiral gauge symmetry, which becomes strongly coupled and hence forms
condensates that self-break the ETC symmetry.  In reasonably
ultraviolet-complete ETC models \cite{at94} it is also necessary to include
another auxiliary, strongly coupled gauge interaction.  Accounting for the
large mass splitting between the $t$ and $b$ quarks may require additional
mechanisms \cite{topcolor} (recent reviews of TC/ETC include
\cite{sanrev,pdg,dewsb}).  In this paper we do not try to construct
quasi-realistic models of dynamical electroweak symmetry breaking but instead
focus on the SU($N$) vectorial gauge theory with fermions in different
representations as an interesting problem in abstract nonperturbative field
theory.

It should be mentioned that, in principle, an asymptotically free, vectorial
gauge theory with a certain set of massless fermions might confine without
producing any spontaneous chiral symmetry breaking.  The spectrum would thus
include a set of massless gauge-singlet composite fermions.  A necessary (but
not sufficient) condition for this to occur is that there should be a matching
of the global chiral anomalies between the fermion fields in the Lagrangian and
the gauge-singlet massless composite fermions \cite{thooft79}.  In our present
study we will focus on the situation in which, as suggested by the analysis of
the Dyson-Schwinger equation for the fermion propagator(s)s, there is
spontaneous chiral symmetry breaking.  In this context, we recall a simple
heuristic physical argument that confinement produces S$\chi$SB, namely that as
a massless fermion heading outward from the interior of a gauge-singlet state
is ``reflected'' back at the boundary, its chirality flips, and this is
equivalent to the presence of a mass term in the effective Lagrangian
\cite{casher}.  However, although our analysis is restricted to
non-supersymmetric gauge theories, we note for completeness that supersymmetric
SU($N$) gauge theories can, for a certain range in the number of chiral
superfields, exhibit confinement without S$\chi$SB \cite{seiberg}.

\subsection{$\beta DS$ Method for Determining Chiral Phase Boundary}

Here we recall a method to estimate the critical value, $N_{R,cr}$ of the
number of fermions in a single representation $R$ beyond which the theory goes
from a phase with spontaneous chiral symmetry breaking to a phase without such
breaking \cite{bds}. The method combines an analysis
of the beta function and coupling constant evolution into the infrared with an
expression for the critical coupling from an approximation solution of the DS
equation, and hence we call it the $\beta DS$ method.

Let us first consider the theory with $N_R$ fermions transforming according to
a single representation $R$.  If $N_R$ is sufficiently small that $b_2 > 0$,
then as the reference scale $\mu$ decreases from large values, $\alpha(\mu)$
increases until it exceeds the critical value $\alpha_{R,cr}$ at which there is
the formation of a bilinear condensate of the fermions
\beq
\langle \bar \psi \psi \rangle \equiv \sum_{j=1}^{{\rm dim}(R_j)} 
\langle \bar\psi_j \psi_j \rangle = \sum_{j=1}^{{\rm dim}(R_j)} 
\langle \bar\psi_{j,L} \psi_{j,R} \rangle + h.c.  
\label{psibarpsi}
\eeq
(For the gauge group SU(2), the condensate can be written in terms of a product
of same-chirality fermions, as discussed below.) If $N_R$ is sufficiently large
that $b_2 < 0$, then the two-loop beta function has an infrared zero at
$\alpha_{IR}$.  The value of $\alpha_{IR}$ is a monotonically decreasing
function of $N_R$, with partial derivative
\beq
\frac{\partial \alpha_{IR}}{\partial N_R} = -\frac{12\pi \, T(R) \, C_2(G) 
[7C_2(G) + 11 C_2(R) ]}{[17C_2(G)^2 - 2 N_R\{5C_2(G)+3C_2(R)\} ]^2}
\label{dadnrform}
\eeq
If the theory only has one type of fermion representation $R$, then as
$N_R$ increases through a critical value $N_{R,cr}$, and the value of
$\alpha_{IR}$ decreases through the critical value $\alpha_{R,cr}$, the
condensate vanishes and the theory goes over to one without any spontaneous
breaking of chiral symmetry.  Setting
\beq
\alpha_{IR}=\alpha_{R,cr}
\label{alfeq}
\eeq 
yields a solution for the critical number $N_{R,cr}$ for this case where the
theory has fermions in only one representation, $R$.  Stated in other terms, if
$N_R < N_{R,cr}$, then as the theory evolves into the infrared, $\alpha(\mu)$
eventually increases above the critical value $\alpha_{R,cr}$, the fermions
condense and gain dynamical masses of order the condensation scale, and the
evolution further into the infrared of the low-energy effective theory
applicable below this scale is governed by a different beta function.  Thus, as
noted above, in this case, $\alpha_{IR}$ is only an approximate infrared fixed
point.  Here, with fermions in a single representation, below the condensation
scale, the beta function would be that of the pure gauge theory with no
fermions, and hence would not have a perturbative infrared fixed point.  The
only light degrees of freedom in this theory would be the Nambu-Goldstone
bosons (NGB's) resulting from the breaking of the global chiral symmetry by the
fermion condensates, and these, being derivatively coupled, become
non-interacting as the energy scale goes to zero.  If, on the other hand, $N_R
> N_{R,cr}$, then $\alpha_{IR} < \alpha_{R,cr}$, so that no condensates form,
there is thus no spontaneous chiral symmetry breaking, and $\alpha_{IR}$ is an
exact infrared fixed point.  As $N_R$ increases to $N_{R,max}$ so that $b_1$
decreases to zero, the value of $b_2$ approaches a nonzero value, so that
$\alpha_{IR} \to 0$.  The value of $b_2$ at $N_R=N_{R,max}$ is
$N(7N+11C_2(R))$.

We next consider the general case of massless fermions transforming according
to several different types of representations, denoted, as above, by the set of
numbers $\{ N_R \}$.  As the reference scale $\mu$ decreases from large values
where the coupling $\alpha(\mu)$ is small, this coupling increases.  There are
then two possibilities: (i) $b_2 > 0$, so that the two-loop beta function does
not have an infrared zero, and the coupling $\alpha(\mu)$ increases until 
it exceeds the critical value for fermion condensation; (ii) $b_2 < 0$, so that
the two-loop beta function does have an infrared zero, and $\alpha(\mu)$
increases toward this value.  Under category (ii) there are two subcategories,
just as there were for the case of a single type of fermion representation, 
(iia) the numbers $\{N_R\}$ are sufficiently small that $\alpha_{IR}$ is
greater than the critical value for some condensate to form, and (iib) 
the numbers $\{N_R\}$ are sufficiently large so that $\alpha_{IR}$ is less than
the critical value for any condensate to form.  In cases (iia) and (iib),
$\alpha_{IR}$ is an approximate and exact infrared fixed point, respectively. 

Let us assume that the set $\{N_R \}$ is such that either case (i) or case
(iia) holds. Then as the scale $\mu$ decreases from large values, the coupling
$\alpha(\mu)$ increases sufficiently so that there is condensation in the most
attractive channel (MAC).  For a channel in which fermions of representations
$R_1$ and $R_2$ form a condensate transforming as $R_{cond.}$,
\beq
R_1 \times R_2 \to R_{cond.}
\label{rchannel}
\eeq
a measure of the attractiveness is 
\beq
\Delta C_2 = C_2(R_1)+C_2(R_2)-C_2(R_{cond.}) \ . 
\label{deltac2}
\eeq
The maximimization of $\Delta C_2$ implies that in a vectorial gauge theory,
the most attractive channels are always of the form
\beq
R \times \bar R \to 1
\label{rrbar}
\eeq
for various $R$, which preserve the gauge invariance.  Furthermore, for
channels of the form (\ref{rrbar}), $\Delta C_2 = 2C_2(R)$, so that 
the criterion for the critical coupling is, in the one-gluon exchange
approximation to the DS equation, 
\beq
\frac{3 \alpha \Delta C_2}{2\pi} = \frac{3 \alpha C_2}{\pi} = 1 \ , 
\label{alfcrit2}
\eeq
as in Eq. (\ref{alfcrit}). It follows that as the theory evolves from high
scales $\mu$ to lower scales, as $\alpha(\mu)$ increases, if it exceeds a
critical value for condensation, the one-gluon exchange approximation predicts
that this will occur first in the channel (\ref{rrbar}) with the largest value
of $C_2(R)$.  Let us denote the scale where this occurs as $\Lambda_1$.  That
is, with this one-gluon approximation to the Dyson-Schwinger equation, the
fermion with the largest value of $C_2(R)$ has the smallest value of
$\alpha_{R,cr}$ and hence forms a condensate at this highest condensation
scale. Associated with this condensation, the fermions transforming according
to this representation gain a dynamical mass of order $\Lambda_1$.  In the
low-energy effective field theory that is applicable at scales below
$\Lambda_1$, these fermions are then integrated out, and the theory evolves in
a manner determined by a new beta function, calculated without these fermions.

For sets of numbers $\{N_R \}$ for which case (iia) holds, we again find that
the partial derivative of $\alpha_{IR}$ with respect to
one of the numbers $N_R$, denoted $N_{R_i}$, with the others, 
$N_{R_j}$ with $j \ne i$, held fixed, is negative:
\beq
\frac{\partial \alpha_{IR}}{\partial N_{R_i}} < 0 \ . 
\label{dadnr_ineq}
\eeq
Hence, the same logic applies as before.  We can start with a set of fermions
$\{N_R \}$ which is such that $b_2 < 0$, so that there is an infrared zero of
the two-loop beta function, and the numbers $N_R$ are sufficiently small that
$\alpha_{IR}$ is large, and the theory forms chiral-symmetry breaking
condensates. We can then increase one of the numbers, $N_{R_i}$, with the
others held fixed.  As we do this, $\alpha_{IR}$ decreases, and eventually
decreases through the critical value given in Eq. (\ref{alfcrit}) for
condensation in the channel $R_i \times \bar R_i \to 1$, at which point this
condensate vanishes.  The condition in Eq. (\ref{alfeq}) then defines a
critical value $N_{R_i,crit}$.  However, in contrast to the simpler case of the
theory with fermions in only a single representation, now the crtical value
$N_{R_i,cr}$ for a given $R_i$ depends on the values of the numbers, $N_{R_j}$,
$j \ne i$, of fermions transforming according to other representations of the
gauge group.  Another tool that has been applied to analyze chiral symmetry
breaking is a conjectured inequality concerning thermal degrees of freedom
\cite{acs,acss}.

\subsection{Global Chiral Symmetry}

For a vectorial SU($N$) theory with $N \ne 2$ with massless fermions in a set
of representations $\{ R\} \equiv \{R_1, R_2,...R_k\}$, such that the numbers
of (Dirac) fermions are $\{N_R\} \equiv N_{R_1}, N_{R_2},...,N_{R_k} \}$, the
formal (classical) global chiral symmetry is $\prod_{i=1}^k {\rm U}(N_{R_i})_L
\times {\rm U}(N_{R_i})_R$. For each $R_i$, the group ${\rm U}(N_{R_i})_L
\times {\rm U}(N_{R_i})_R$ can be rewritten as
\beq
{\rm SU}(N_{R_i})_L \times {\rm SU}(N_{R_i})_R \times 
{\rm U}(1)_{R_i,V} \times {\rm U}(1)_{R_i,A} \ . 
\label{formalgroup}
\eeq
The vectorial global symmetry U(1)$_{R_i,V}$ represents the conservation of
fermion number for the fermions in the representation $R_i$.  Each of the $k$
axial global symmetries ${\rm U}(1)_{R_i,A}$ is broken by SU($N$) instantons
\cite{thooft}, with divergences of the corresponding axial-vector currents
$\partial_\lambda J^{A,\lambda}_{R_i} \propto
[\alpha/(4\pi)]T(R_i)F^a_{\mu\nu}\tilde F^{a,\mu\nu}$.  From these $k$ broken
symmetries ${\rm U}(1)_{R_i,A}$, $i=1,...,k$, one can construct $k-1$ linear
combinations that are conserved in the presence of instantons, which we denote
${\cal U}(1)_{s,A}$, $s=1,...,k-1$ with currents ${\cal J}^{A \lambda}_s$.  Let
us define
\beq
\hat J^{A,\lambda}_{R_i} \equiv \frac{J^{A,\lambda}_{R_i}}{T(R_i)} \ . 
\label{jahat}
\eeq
One of the conserved currents is (up to a normalization factor) 
\beq
{\cal J}^{A \lambda}_1 \propto {\hat J}^{A,\lambda}_{R_1} - 
{\hat J}^{A,\lambda}_{R_2} \ . 
\label{ja1}
\eeq
The others are constructed by Gram-Schmidt orthonormalization.  For
example, for $k=3$, the other one is 
\beq
{\cal J}^{A \lambda}_2 \propto \frac{1}{2}
\left [ {\hat J}^{A,\lambda}_{R_1} + {\hat J}^{A,\lambda}_{R_2} - 
2{\hat J}^{A,\lambda}_{R_3} \right ] \ . 
\label{ja2keq3}
\eeq
Thus the actual (continuous) nonanomalous global symmetry of the theory, before
any fermion condensates form, is
\beqs
G_{global} & = & \left [\prod_{i=1}^k {\rm SU}(N_{R_i})_L \times 
{\rm SU}(N_{R_i})_R \times {\rm U}(1)_{R_i,V} \right ] \cr\cr 
& \times & \left [ \prod_{s=1}^{k-1} {\cal U}(1)_{A,s} \right ] \ . 
\label{globalgroup}
\eeqs

The resultant realization of this global symmetry depends on the gauge coupling
evolution and whether the coupling $\alpha$ increases above the critical value
for condensation of the fermions in the $R_i$ representation.  As an example,
let us assume that all of the fermions condense, at the respective different
scales $\Lambda_{N_{R_i}}$, $i=1,...,k$. For our discussion here we shall label
the representation with the largest value of $C_2(R_i)$ as $R_1$; from the
one-gluon exchange approximation, it then follows that the $R_1$ fermions
condense at the highest scale, $\Lambda_1$.  In
accordance with the most-attractive channel arguments recalled above, the
fermion condensate of the form $\langle \bar\psi_{R_1} \psi_{R_1}\rangle$
preserves the global U(1)$_{R_1,V}$ and breaks the non-Abelian global symmetry
from ${\rm SU}(N_{R_1})_L \times {\rm SU}(N_{R_1})_R$ to its diagonal,
vectorial subgroup, ${\rm SU}(N_{R_1})_V$. This condensate also
breaks each of the $k-1$ ${\cal U}(1)_{s,A}$ axial symmetries.  In the
low-energy effective field theory applicable at scales $\mu < \Lambda_1$,
with the fermions in the $R_1$ representation having gained dynamical masses of
order $\Lambda_1$ and having been integrated out, one can construct $k-2$
appropriate linear combinations of the former $k-1$ ${\cal U}(1)_{s,A}$ axial
symmetries that exclude the $R_1$ fermions and are preserved
in the presence of instantons.  We denote these as ${\cal U}(1)'_{s,A}$.  The
continuous global symmetry group of this low-energy effective theory below
$\Lambda_{R_1}$ is then
\begin{widetext}
\beq
G_{global}' = \left [\prod_{i=2}^k {\rm SU}(N_{R_i})_L \times 
{\rm SU}(N_{R_i})_R \right ] \times 
\left [\prod_{i=1}^k {\rm U}(1)_{R_i,V} \right ] \times 
\left [ \prod_{s=1}^{k-2} {\cal U}(1)'_{A,s} \right ] \ . 
\label{globalgroup2}
\eeq
\end{widetext}
The number of broken generators of continuous global Lie algebras at the first
scale is $N_{R_1}^2-1$ from the breaking of the non-Abelian group, plus one for
the breaking of one linear combination of the $k-1$ nonanomalous axial U(1)
symmetries, for a total of $N_{NGB, \Lambda_{R_1}} = N_{R_1}^2$
Nambu-Goldstone bosons resulting from this first level of fermion
condensation.  One repeats this process at each of the various condensation
scales.  The NGB's produced at each level couple derivatively, and hence become
progressively more weakly interacting as powers of $\mu/f_{R_i}$, where
$f_{R_i}$ is the generalization of the pion decay constant applicable to the
condensation of the $R_i$ fermions. 

In the case $N=2$, because SU(2) has only (pseudo)real representations, the
analysis of the global symmetry is different than in the case of SU($N$) with
$N \ne 2$.  If, for example, one has an SU(2) theory with $N_f$ (Dirac)
fermions in the fundamental representation, then one can reexpress these
fermions as a set of $2N_f$ chiral (say, left-handed) fermions, and the
covariant derivative term has the form $\bar \psi_L \gamma \cdot D \psi_L$,
where $\psi$ is a $2N_f$-dimensional vector of left-handed fermions.  It
follows that the formal (classical) global symmetry in this case is ${\rm
U}(2N_f)_L$, or equivalently, ${\rm SU}(2N_f)_L \times {\rm U}(1)_L$. The
U(1)$_L$ is broken by the SU(2) instantons \cite{thooft}, so that the
nonanomalous global symmetry is ${\rm SU}(2N_f)_L$. The condensates are of the
form $\langle \epsilon_{ab}\psi^{a \ T}_{p,L} C \psi^b_{p',L} \rangle$, where
$\epsilon_{ab}$ is the antisymmetric tensor density for SU(2) and $1 \le p, \
p' \le 2N_f$.  If the fermions are in the rank-2 symmetric (equivalently, the
adjoint) representation, of the form $\psi^{ab}_{p,L}$ with $1 \le p \le 2N_f$,
then the condensate are of the form $\langle \epsilon_{ar}\epsilon_{bs}\psi^{ab
\ T}_{p,L} C \psi^{rs}_{p',L} \rangle$, and so forth for higher-dimensional
representations.  These condensates break the ${\rm SU}(2N_f)_L$ down to its
symplectic subgroup, ${\rm Sp}(2N_f)_L$.  In this case there are thus $N_{NGB}
= 2N_f^2-N_f-1$ Nambu-Goldstone bosons.

\section{SU($N$) Gauge Theory with Fermions in a Single Representation}

In this section we review some results on an SU($N$) gauge theory with fermions
in a single representation, which will serve as a useful background for our
analysis of the theory with fermions in multiple different representations. 

\subsection{Fundamental Representation}

For the SU($N$) theory with $N_F$ Dirac fermions in the fundamental
representation $F$ ($=\fund$ in Young tableau notation), the condition for
asymptotic freedom yields the upper bound $N_F < N_{F,max}$, where
\beq
N_{F,max} = \frac{11N}{2} \ . 
\label{nfmax_fund}
\eeq
The coefficient $b_2$ changes sign from positive to negative as $N_F$ increases
through the value
\beq
N_{F,IR} = \frac{34N^3}{13N^2-3} \ , 
\label{nfir_fund}
\eeq
which is always less than $N_{F,max}$.  For $N_{F,IR} < N_F < N_{F,max}$, the
beta function has a zero away from the origin at
\beq
\alpha_{IR} = \frac{4\pi(11N-2N_F)}{-34N^2+N_F(13N-3N^{-1})} \ . 
\label{alfir_fund}
\eeq
The estimate for the critical value for condensation (from Eq. (\ref{alfcrit})
is
\beq
\alpha_{F,cr} = \frac{2\pi N}{3(N^2-1)} \ . 
\label{alfcrit_fund}
\eeq
Setting $\alpha_{IR} = \alpha_{F,cr}$ and solving for $N_F$, one obtains the 
critical value of $N_F$ \cite{bds} 
\beq 
N_{F,cr} = \frac{2N(50N^2-33)}{5(5N^2-3)} \ . 
\label{nfcr_fund}
\eeq
As $N \to \infty$, this has the series expansion 
\beq
N_{F,cr} = N \left [ 1 - \frac{3}{50N^2} - \frac{9}{250N^4} - 
O \left ( \frac{1}{N^6} \right ) \right ] \ . 
\label{nfcr_fund_largen}
\eeq
For $N=2$, $N_{F,cr} \simeq 8$ and for $N=3$, $N_{F,cr} \simeq 12$. Recent
lattice measurements for the $N=3$ case are in broad agreement, to within the
uncertainties, with this prediction \cite{afn}. 

The DS equation analysis is semi-perturbative in the sense that it contains
polynomial dependence on $\alpha$, and it neglects nonperturbative effects
associated with confinement and instantons.  The DS equation is an integral
equation, and the standard analysis of this equation involves an integration
over Euclidean loop momentum $k$ from $k=0$ to $k=\infty$.  If the theory
confines, then the lower bound for the Euclidean loop momentum should actually
not be $k=0$, but instead $k = k_{min.} \sim r_c^{-1}$ where $r_c$ is the
spatial confinement scale \cite{lmax}.  The use of $k=0$ thus overestimates the
tendency toward S$\chi$SB.  Instantons enhance S$\chi$SB, and the neglect of
instanton effects amounts to an underestimate of the tendency toward S$\chi$SB;
since these two neglected aspects of the physics - confinement and instantons -
produce errors that are of opposite sign as regards the tendency for S$\chi$SB,
it is plausible that these errors tend to cancel out, so this may help to
explain why the usual DS analysis may be reasonably accurate \cite{lmax}, at
least in the case $N=3$ where recent lattice results are broadly consistent
with it.

\subsection{Rank-2 Symmetric and Antisymmetric Representations} 

In this section we consider the two separate cases of the SU($N$) theory
with (i) $N_S$ fermions in the symmetric rank-2 representation, $S \equiv \sym$
and (ii) $N_A$ fermions in the antisymmetric rank-2 representation, $A \equiv
\asym$.  Since a number of formulas are similar for these two cases, we include
themtogether in this section.  In the case of $R=\sym$, our analysis applies
for any $N \ge 2$, while for $R=\asym$, we take $N \ge 4$, since for $N=2$,
$\asym$ is the singlet and for $N=3$, $\asym$ is not a distinct representation,
but is instead equivalent to $\overline{\fund}$.).  For the SU($N$) theory with
$N_s$ Dirac fermions in the symmetric rank-2 representation $\sym$, the
condition for asymptotic freedom yields the upper bounds $N_S < N_{S,max}$,
where
\beq
N_{S,max} = \frac{11N}{2(N+2)}
\label{nfmax_s}
\eeq
and $N_A < N_{A,max}$, where
\beq
N_{A,max} = \frac{11N}{2(N-2)} \ . 
\label{nfmax_a}
\eeq
The coefficient $b_2$ changes sign from positive to negative 
as $N_S$ and $N_A$ increase through the respective values
\beq
N_{S,IR} = \frac{17N^3}{(N + 2)(8N^2 + 3N - 6)}
\label{nsir}
\eeq
and
\beq
N_{A,IR} = \frac{17N^3}{(N - 2)(8N^2 - 3N - 6)} \ , 
\label{nair}
\eeq
which are always less than the respective values $N_{S,max}$ and 
$N_{A,max}$. 

For the theory with just $N_S$ fermions in the $S$ representation, and
$N_{S,IR} < N_S < N_{S,max}$, the beta function (\ref{beta}) has a zero away
from the origin at
\beq
\alpha_{IR,S} = \frac{2\pi(11N-2N_S(N+2))}{-17N^2+N_S(8N^2+19N-12N^{-1})} \ . 
\label{alfir_s}
\eeq
The estimate for the critical value for condensation (from Eq. (\ref{alfcrit})
is
\beq
\alpha_{S,cr} = \frac{\pi N}{3(N+2)(N-1))} \ . 
\label{alfcrit_s}
\eeq
Setting $\alpha_{IR,S} = \alpha_{S,cr}$ and solving for $N_S$, we obtain the
critical value of $N_S$,
\beq
N_{S,cr} = \frac{N(83N^2+66N-132)}{5(N+2)(4N^2+3N-6)}  \ . 
\label{nscr}
\eeq
For $N \to \infty$, this has the series expansion
\beq
N_{S,cr} = \frac{83}{20} - \frac{649}{80N} + \frac{5027}{320N^2} 
+ O \left ( \frac{1}{N^3} \right ) \ . 
\label{nscrlargen}
\eeq
For $N=2$, $N_{S,cr} \simeq 2.1$, while for $N=3$, $N_{S,cr} \simeq 2.5$.  Some
Lattice measurements for the $N=3$ case are reported in \cite{lgtsym}. As $N$
increases from 2 to $\infty$, $N_{S,cr}$ increases monotonically from $83/40
\simeq 2.08$ to $83/20 = 4.15$. (As before, although we quote the exact
fractions and give the floating-point numbers to three significant figures, we
emphasize that because of the strong-coupling nature of the physics and the
approximations involved, these numbers have estimated theoretical uncertainties
of O(1). This applies to all such estimates of $N_{R,cr}$ values in this
paper.)

For the theory with just $N_A$ fermions in the $A$ representation, and
$N_{A,IR} < N_A < N_{A,max}$, the beta function (\ref{beta}) has a zero away
from the origin at
\beq
\alpha_{IR,A} = \frac{2\pi (11N-2N_A(N-2))}{-17N^2+N_A(8N^2-19N+12N^{-1})} \ . 
\label{alfir_a}
\eeq
The estimate for the critical value for condensation (from Eq. (\ref{alfcrit})
is
\beq
\alpha_{A,cr} = \frac{\pi N}{3(N-2)(N+1))} \ . 
\label{alfcrit_a}
\eeq
Setting $\alpha_{IR,A} = \alpha_{A,cr}$ and solving for $N_A$, we obtain the
critical value
\beq
N_{A,cr} = \frac{N(83N^2-66N-132)}{5(N-2)(4N^2-3N-6)} \ . 
\label{nacr}
\eeq
For $N \to \infty$, this has the series expansion
\beq
N_{A,cr} = \frac{83}{20} + \frac{649}{80N} + \frac{5027}{320N^2} 
+ O \left ( \frac{1}{N^3} \right )  \ . 
\label{nacrlargen}
\eeq
As $N$ increases from 3 to $\infty$, $N_{A,cr}$ decreases monotonically from
$417/35 \simeq 11.9$ to $83/20 \simeq 4.15$.

\subsection{Adjoint Representation}

For the case of $N_{Adj}$ Dirac fermions, or equivalently, $2N_{Adj,Maj}$
Majorana fermions, in the adjoint representation $Adj$,
the condition for asymptotic freedom is $N_{Adj} < N_{Adj,max}$, where
\beq
N_{Adj,max} = \frac{11}{4} \ , 
\label{nfmax_adj}
\eeq
i.e., $N_{Adj} \le 2$.  Majorana fermions in the adjoint representation of the
gauge group appear naturally in supersymmetric theories.  In the present
non-supersymmetric context, we shall restrict ourselves to adjoint fermions of
Dirac type.  The coefficient $b_2$ changes sign from positive to 
negative as $N_{Adj}$ increases through the value
\beq
N_{Adj,IR} = \frac{17}{16} \ . 
\label{nfir_adj}
\eeq
For $N_{Adj,IR} < N_{Adj} < N_{Adj,max}$, the
beta function has a zero away from the origin at
\beq
\alpha_{IR} = \frac{2\pi(11-4N_{Adj})}{N(-17N+16N_{Adj})} \ . 
\label{alfir_adj}
\eeq
Setting 
\beq
\alpha_{Adj,cr} = \frac{\pi}{3N}
\label{alfcrit_adj}
\eeq
equal to $\alpha_{Adj,cr}$, one solves for 
\beq 
N_{Adj,cr} = \frac{83}{40} = 2.075 \ . 
\label{nfcr_adj}
\eeq

\section{SU(2) Gauge Group}

For the simplest non-Abelian Yang-Mills gauge group, SU(2), we can give a
rather compact general treatment that includes all possible representations.
We recall that this group has only (pseudo)-real representations $R$, which are
labeled by a single Dynkin index, the non-negative integer $p_1=2I$, where $I$
will be labelled as the ``isospin'' (not to be confused with the actual gauged
weak isospin). $I=1/2$ is the fundamental representation, $\fund$;
$I=1$ is the adjoint or equivalently, rank-2 symmetric representation, $\sym$;
$I=3/2$ is the rank-3 symmetric representation, and so forth.  The following
SU(2) relations will be useful:
\beq
C_2(I) = I(I+1)
\label{c2su2}
\eeq
and
\beq
T(I) = \frac{(2I+1) \, I(I+1)}{3} \ . 
\label{tsu2}
\eeq
The asymptotic freedom condition (\ref{afcondition}) reads 
\beq
\sum_I N_I T(I) < \frac{11}{2} \ , 
\label{b1pos_su2}
\eeq
where the sum over $I$ is formally over all positive integral and half-integral
values, but actually truncates, because of fact that
$C_2(I) > 11/2$ for $I \ge 2$.  Hence, (\ref{b1pos_su2}) reduces to the
Diophantine inequality 
\beq
\frac{1}{2}N_{1/2} + 2N_{1} + 5N_{3/2} < \frac{11}{2} \ . 
\label{b1pos_su2_explicit}
\eeq
The nontrivial solutions to this include cases with only one type
of fermion representation present.  In these cases, the allowed numbers of
fermions of each type are 
\beq
N_{1/2} \le 10 
\label{su2sol_half}
\eeq
\beq
N_1 \le \left [\frac{11}{4} \right ]_\ell = 2
\label{su2sol_adj}
\eeq
and
\beq
N_{3/2} \le \left [\frac{11}{10} \right ]_\ell = 1
\label{su2sol_32}
\eeq
where here $[\nu]_\ell$ denotes the greatest integer less than or equal to 
$\nu$ and it is understood in each case that the $N_I$'s for other $I$'s are
zero.  We also find the following solutions of the asymptotic freedom condition
with two different fermion representations present (and $N_{3/2}=0$):
\beq
1 \le N_{1/2} \le 6,  \quad N_1=1 
\label{mixsol1}
\eeq
and
\beq
 1 \le N_{1/2} \le 2, \quad N_1=2 \ . 
\label{mixedsol2} 
\eeq

Substituting the general result for $C_2(I)$ in Eq. (\ref{c2su2}) in Eq. 
(\ref{alfcrit}), we have, in this approximation, 
\beq
\alpha_{I,cr} =  \frac{\pi}{3I(I+1)} \ . 
\label{alfcritsu2}
\eeq

The predictions for the case of $N_{1/2}$ massless Dirac fermions in the
fundamental representation are well known \cite{bds}.  The two-loop coefficient
$b_2$ reverses sign from positive to negative as $N_{1/2}$ increases through
the value $272/49 \simeq 5.55$ and decreases through negative values as
$N_{1/2}$ increases.  The zero of the two-loop beta function occurs at
\beq
\alpha_{IR} = \frac{16\pi(11-N_{1/2})}{49N_{1/2}-272} \quad SU(2), \quad I=1/2
\ . 
\label{alfir_su2_fund}
\eeq
Equating $\alpha_{IR} = \alpha_{1/2,cr}=4\pi/9$ or substituting $N_c=2$ into
Eq. (\ref{nfcr_fund}), one obtains the critical value for the case with only
fermions in the $I=1/2$ representation, $N_{1/2,cr} = 668/85 \simeq 7.9$.

There are two other cases where the theory involves only fermions of a single
type of representation, namely those with $I=1$ and $I=3/2$.  For the symmetric
rank-2 tensor, or equivalently adjoint, representation, $I=1$, substituting
$N_c=2$ into Eq. (\ref{nsir}) or using (\ref{nfir_adj}) shows that $b_2$
reverses sign from positive to negative as $N_1$ increases through the value
17/16.  Similarly, substituting $N_c=2$ into Eq. (\ref{alfir_s}) or using 
(\ref{alfir_adj}), one derives that
\beq
\alpha_{IR, \ I=1} = \frac{\pi(11-4N_1)}{16N_1-17} \ . 
\label{alfir_su2_adj}
\eeq
Setting this equal to $\alpha_{cr,I=1}=\frac{\pi}{6} \simeq 0.52$ or using 
Eq. (\ref{nfcr_adj}) directly, one has 
$N_{cr, \ I=1} = 83/40 = 2.075$. 

Finally, among the cases with a single fermion representation present, there is
the case of fermions with $I=3/2$.  For this case, $b_2$ reverses sign from
positive to negative as $N_{3/2}$ increases through the value $N_{3/2} =
8/25 = 0.32$.  The two-loop beta function has a zero away from the origin
at
\beq
\alpha_{IR, \ I=3/2} = \frac{8 \pi(11-10N_{3/2})}{17(25N_{3/2}-8)} \ . 
\label{alfir_su2_32}
\eeq
Setting this equal to $\alpha_{cr,\ I=3/2}= 4\pi/45 \simeq 0.28$, we get 
the critical value 
\beq
N_{cr, \ I=3/2} = \frac{1126}{1325} \simeq 0.85 \ . 
\label{nfcrit_su2_32}
\eeq
Since $\alpha_{IR}$ decreases with increasing $N_{3/2}$ and since the
minimal nonzero value is $N_{f,cr, \ I=3/2}$ is 1, this predicts that with one
such Dirac fermion with $I=3/2$, the infrared fixed point is below the value
for condensation and hence is an exact IR fixed point.  That is, the gauge 
coupling will evolve to this point without any condensate involving the 
$I=3/2$ forming, so that it does not gain any dynamical mass and remains
massless. Thus, in the infrared limit of this theory the fermion is massless.
We summarize our results for SU(2) in Table \ref{su2table}.

\subsection{SU(2) Theory with Fermions in Several Representations}

We next consider the SU(2) theory with fermions in several different 
representations.  As discussed above, the requirement of asymptotic 
freedom limits the possible numbers of fermions, delineated by the numbers 
$N_{1/2}$, $N_1$, and $N_{3/2}$.  For the case of $N_{3/2}=0$, we have
\beq
b_2 = \frac{1}{3} \left [ 136 - \frac{49N_{1/2}}{2} - 128N_1 \right ] \ . 
\label{b2halfone}
\eeq
For $N_1=1$ and $N_{1/2}=0, \ 1$, it follows that $b_2 > 0$, so that the beta
function has no infrared zero away from the origin.  This means that as the
scale $\mu$ decreases, $\alpha$ increases until it exceeds the value
$\alpha_{cr,I=1}$, and the $I=1$ fermions condense.  They are then integrated
out, and the theory evolves further into the infrared as governed by the beta
function with only the $I=1/2$ fermions present.  The coupling thus increases
further until it exceeds the value $\alpha_{cr,I=1/2}$, at which point these
$I=1/2$ fermions condense.

For $N_1=1$ and $1 \le N_{1/2} \le 6$, $b_2 < 0$ and so the beta
function has an infrared zero away from the origin, at
\beq
\alpha_{IR} = \frac{16\pi(11-N_{1/2}-4N_1)}{(49N_{1/2}+256N_1-272)} \ . 
\label{alpha_ir_su2halfadj}
\eeq
If this is less than the critical value (\ref{alfcrit}), then no
fermion condensates form.  Setting this $\alpha_{IR}$ equal to the smaller of
the two critical values, $\alpha_{cr,I=1}$, one derives the condition for
condensation of the $I=1$ fermions.  This is 
\beq
N_{1/2}+\frac{128}{29}N_1 < \frac{1328}{145} \simeq 9.16 \ . 
\label{ineqsu2}
\eeq
In addition to the cases with $N_1=0$ dealt with above, this condition is
satisfied for $N_1=1$ and $1 \le N_{1/2} \le 4$.  For $N_1=1$ and $N_{1/2}=5$,
$\alpha_{IR} = 0.44$, which is close enough to $\alpha_{cr}=\pi/6 = 0.52$ so
that, given the uncertainties in the calculation, there might or might not be
condensation of the $I=1$ fermions. For $N=1$ and $N_{1/2}=6$, $\alpha_{IR} =
0.18$, which is below the value for condensation of both the $I=1$ and $I=1/2$
fermions.  Hence, in this case, this is an exact infrared fixed point, and
the theory evolves into the infrared without any spontaneous chiral symmetry
breaking.

For $N_1=2$, the condition of asymptotic freedom, $2N_{1/2}+8N_1 < 22$, is
$N_{1/2} < 3$. Aside from the case $N_{1/2}=0$ dealt with above, for the cases
$N_{1/2}=1$ and $N_{1/2}=2$, the two-loop beta function has an infrared zero at
the respective values $\alpha = 32 \pi/289 \simeq 0.35$ and $\alpha = 8 \pi/169
\simeq 0.15$, both of which are smaller than the estimate $\alpha_{cr,I=1} =
\pi/6$, so that the $\beta DS$ analysis predicts that no condensate occurs and
the theory evolves into the infrared in a phase without any spontaneous chiral
symmetry breaking.

\section{SU(3) Gauge Group}

It is also of interest to investigate properties of a (vectorial,
asymptotically free) SU(3) gauge theory with multiple fermion representations. 
We recall that the representations of SU(3) are labelled by a set of two 
Dynkin indices $(p_1,p_2)$, where $p_i$ are non-negative integers.  We use the
following results from group theory.  The dimension of the representation is
\beq
{\rm dim}(p_1,p_2) \equiv d(p_1,p_2) = 
(1+p_1)(1+p_2)\left ( 1 + \frac{p_1+p_2}{2} \right )  \ . 
\label{dr}
\eeq
The quadratic Casimir invariant is
\beq
C_2(p_1,p_2) = \frac{1}{3} \bigg [ p_1^2+p_2^2+p_1p_2+3(p_1+p_2) \bigg ]
\label{c2r}
\eeq
and the trace invariant is 
\beq
T(p_1,p_2) = \frac{{\rm dim}(p_1,p_2) \, C_2(p_1,p_2)}{8} \ . 
\label{tr}
\eeq
The asymptotic freedom condition (\ref{afcondition}) reads 
\beq
\sum_R N_R T(R) < \frac{33}{4} \ , 
\label{b1pos_su3}
\eeq
where, again, the sum over representations truncates because for sufficiently
large values of $p_1$ and/or $p_2$, $T(p_1,p_2) > 33/4$.  We find that it is
satisfied by the following nonsinglet representations labelled by their
dimension and values of $(p_1,p_2)$:
\beq
R_{(p_1,p_2)} = 3_{(1,0)} \ \ 6_{(2,0)}, \ 8_{(1,1)}, \ 10_{(3,0)} \ . 
\label{su3reps}
\eeq
The asymptotic freedom condition (\ref{b1pos_su3}) thus can be written
explicitly as the Diophantine inequality 
\beq
\frac{1}{2}N_3 + \frac{5}{2}N_6 + 3N_8 + \frac{15}{2}N_{10} < \frac{33}{4} \ . 
\label{b1pos_su3_explicit}
\eeq
In the case where the theory has fermions in only one of these representations
$R=(p_1,p_2)$, the upper bounds on the corresponding number $N_R$ are, in
addition to $N_3 \le [33/2]_\ell=16$, 
\beq
N_6 \le \left [\frac{33}{10} \right ]_\ell = 3
\label{n6upper}
\eeq
\beq
N_8 \le \left [\frac{11}{4} \right ]_\ell = 2
\label{n8upper}
\eeq
and
\beq
N_{10} \le \left [\frac{11}{10} \right ]_\ell = 1 \ . 
\label{n10upper}
\eeq
In each of these inequalities, it is understood that the $N_R$'s for other
representations are zero. 

For the case of multiple fermion representations, we find that the asymptotic
freedom condition is satisfied for the following combinations of two
fermion representations (where $N_R$'s that do not appear are zero):
\beq
1 \le N_3 \le 11, \quad N_6=1
\label{comb1}
\eeq
\beq
1 \le N_3 \le 6, \quad N_6=2
\label{comb2}
\eeq
\beq
N_3 =1, \quad N_6=3
\label{comb3}
\eeq
\beq
1 \le N_3 \le 10, \quad N_8 = 1
\label{comb4}
\eeq
\beq
1 \le N_3 \le 4, \quad N_8 = 2
\label{comb5}
\eeq
\beq
1 \le N_6 \le 2, \quad N_8=1
\label{comb6}
\eeq
and
\beq
N_3=1, \quad N_{10}=1 \ . 
\label{comb7}
\eeq
We also find the asymptotic freedom condition allows the following 
combination of three fermion representations:
\beq
1 \le N_3 \le 5, \quad N_6=1, \quad N_8=1 \ . 
\label{comb8}
\eeq
It is straightforward to calculate the values of $b_2$ for each of the various
sets $\{N_R\}$ involving one or several different fermion representations.  As
before, if $b_2 > 0$, then $\alpha$ definitely increases past $\alpha_{R,cr}$
for at least one of the fermion representations $R$, and one analyzes the
sequential condensations accordingly.  If $b_2 < 0$, then one determines
whether the behavior is of type (iia) or (iib) in the classification discussed
above, i.e. whether $\alpha_{IR}$ is greater than $\alpha_{R,cr}$ for some $R$
or $\alpha_{IR}$ is less than the minimum $\alpha_{R,cr}$.  All of these types
of behavior are exhibited by various sets $\{R\}$ among those allowed by
asymptotic freedom.

\section{Relative Scales of Condensation}

In an SU($N$) gauge theory with $N_R$ fermions in a single representation $R$
and with a small, perturbatively calculable value of $\alpha(\mu_{UV})$ at some
high scale, $\mu_{UV}$, provided that $N_R$ is sufficiently small that there 
exists a scale $\mu=\Lambda_R$ at which $\alpha(\mu)$ increases beyond the
critical value $\alpha_{R,cr}$ for condensation, then one can estimate this
scale by integrating the renormalization group equation, with the leading-order
result
\beq
\Lambda_R \simeq \mu_{UV} \exp \left [ -\frac{2\pi}{b_1(R)} 
\left (\alpha(\mu_{UV})^{-1} - \alpha_{R,cr}^{-1} \right ) \right ] \ , 
\label{lamr}
\eeq
where we have indicated explicitly the dependence of the beta function
coefficient $b_1=(1/3)(11N_c-4N_RT(R))$ on $R$.  One can, of course, calculate
$\Lambda_R$ to greater accuracy by including higher-order terms in the beta
function, as well as estimates of important physics effects not included in the
perturbative beta function, such as instantons, but this leading-order result
will be sufficient for our discussion here. From Eq. (\ref{lamr}), it follows
that if one compares an SU($N$) theory with fermions in the single
representation $R_i$ with a different SU($N$) theory with fermions in the
single representation $R_j$, the ratio of the condensation scales,
$\Lambda_{R_i}/\Lambda_{R_j}$, depends on all of the parameters $\mu_{UV}$,
$N_{R_i}$, and $N_{R_j}$, as well as the ladder estimates for the respective
critical couplings, $\alpha_{R_i,cr}$ and $\alpha_{R_j,cr}$.  For fixed values
of $\mu_{UV}$, $\alpha_{UV}(\mu)$, and $N_{R_i}$, one may ask how
$\Lambda_{R_i}/\mu_{UV}$ depends on $R_i$.  There are two countervailing
effects that are relevant here: (i) as the dimension ${\rm dim}(R_i)$ of a
representation $R_i$ increases, the value of $C_2(R_i)$ also tends to increase
(although the dependence is not necessarily monotonic \cite{c2n}), and hence
the critical value of the coupling, $\alpha_{R_i,cr}$ decreases; if this
increase were the only effect, then $\Lambda_{R_i}/\mu_{UV}$ would increase
with increasing size of $R_i$.  However, there is an effect that goes in the
opposite direction, namely, (ii) as the dimension ${\rm dim}(R_i)$ of the
representation $R_i$ increases, the value of $T(R_i)$ also increases, thereby
reducing $b_1(R_i)$, and slowing down the increase of $\alpha$ as $\mu$
descends from $\mu_{UV}$.  Indeed, a sufficient increase in the size of the
representation $R_i$, for a fixed $N_{R_i}$ can even change the infrared
behavior of the theory to preclude any spontaneous chiral symmetry breaking and
condensate formation.  Thus, for a general $R_i$, one cannot draw a very robust
conclusion about how, for fixed values of $\mu_{UV}$, $\alpha(\mu_{UV})$, and
$N_{R_i}$, the condensation scale $\Lambda_{R_i}$ depends on the size of $R_i$.

In situations in which the theory has fermions in two or more different
representations and these form condensates at different mass scales, it is of
interest to calculate the ratio(s) of these scales. In carrying out this
analysis, one acknowledges that, owing to the fact that the theory is strongly
coupled at these scales, it is only possible to obtain rough estimates of such
a ratio of condensation scales. Let us consider the SU($N$) theory with the
specific set of Dirac fermions $\{N_R\}=\{N_{R_1},N_{R_2}\}$, say, where the
$R_i$, $i=1,2$ are two different (nonsinglet) representations of SU($N$).
Without loss of generality, we label the representations such that $C_2(R_1) >
C_2(R_2)$. As always, we require that this set $\{N_R\}$ have the property that
the theory is asymptotically free, and here we also require that the set is
such that condensates of both types of fermions occur, since otherwise there is
no ratio to estimate.  Again, we assume that at the high reference scale
$\mu_{UV}$ the coupling $\alpha(\mu_{UV})$ is small and the theory is
perturbatively calculable.  As $\mu$ decreases from $\mu_{UV}$, the first
condensation occurs when $\alpha(\mu) = \alpha_{R_1,cr}$, where
$\alpha_{R_1,cr}$ was given in Eq. (\ref{alfcrit}), from the solution of the
Dyson-Schwinger equation in the approximation of one-gauge-boson exchange.
Solving the renormalization group equation to leading order, we have, for the
scale at which this condensation occurs the result
\beqs
\Lambda_1 & \simeq & \mu_{UV} \, \exp \left [ -\frac{2 \pi}{b_1} 
\left ( \alpha(\mu_{UV})^{-1} - \alpha_{R_1,cr}^{-1} \right ) \right ] \cr\cr
          & \simeq & \mu_{UV} \, \exp \left [ -\frac{2 \pi}{b_1} 
\left ( \alpha(\mu_{UV})^{-1} - \frac{3C_2(R_1)}{\pi} \right ) \right ] \ , 
\cr\cr
& & 
\label{lambda_1}
\eeqs
where $b_1$ is given by the appropriate special case of Eq. (\ref{b1}) with the
full set $\{N_{R_1},N_{R_2}\}$ of fermions.  The $N_{R_1}$ fermions in the
condensates gain dynamical masses of order $\Lambda_1$ and are integrated out
of the low-energy effective field theory applicable for $\mu < \Lambda_1$.  The
coupling $\alpha(\mu)$ continues to grow, as governed by the beta function of
this low-energy effective theory, which differs from that of the high-scale
theory by the removal of the $N_{R_1}$ fermions in the representation $R_1$.
Insofar as the coupling $\alpha$ is not too large to prevent one from using the
perturbative beta function to track its evolution reliably for $\mu <
\Lambda_1$, one has
\beq
\alpha(\mu)^{-1} = \alpha^{-1}(\Lambda_1) +
\frac{b_1(R_2)}{2\pi}\ln \left (\frac{\Lambda_1}{\mu} \right ) \ , 
\label{alfinvlow}
\eeq
where $b_1(R_1)$ is the value of $b_1$ from Eq. (\ref{b1}) for the low-energy
effective field theory with only $N_{R_2}$ fermions in $R_2$ present.
Then, given our assumptions about the set $\{N_{R_1},N_{R_2}\}$, at a lower
scale $\Lambda_2$, condensation occurs for the fermions in the representation
$R_2$, when $\alpha(\mu) = \alpha_{R_2,cr}$.  Solving for the ratio of these 
two condensations scales in this rough approximation, we obtain
\beq
\frac{\Lambda_2}{\Lambda_1} \simeq \exp \left [ -\frac{6}{b_1(R_2)} 
\bigg ( C_2(R_1) - C_2(R_1) \bigg ) \right ] \ , 
\label{lamrat}
\eeq
where $b_1(R_2) = (1/3)(11N - 4N_{R_2}T(R_2))$.  As an example, consider the
SU(2) theory with $R_1$ and $R_2$ being the $I=1$ and $I=1/2$ representations,
respectively, and numbers $N_{R_1} \equiv N_1$ and of $N_{R_2} \equiv 
N_{1/2}1$ for which there are two condensations, as indicated in Table 
\ref{su2table}.  Then
\beq
\frac{\Lambda_{1/2}}{\Lambda_1} \simeq
 \exp \left [ -\frac{45}{4(11-N_{1/2})} \right ] \ . 
\label{lamratsu2}
\eeq
As $N_{1/2}$ increases from 2 to 4,  this ratio 
$\Lambda_{1/2}/\Lambda_1$ decreases from about 0.3 to 0.2. These are comparable
to the sort of ratios of condensation scales that would characterize the
sequential breaking of reasonably ultraviolet-complete extended-technicolor 
theories (e.g., \cite{nt,ckm}).

\section{Effects of Nonzero Intrinsic Masses for Fermions}

In the discussion up to this point we have assumed that the fermions have zero
intrinsic masses in the Lagrangian describing the high-scale physics, and the
only masses that they acquire arise dynamically if they are involved in
condensates that form as the gauge interaction becomes sufficiently strongly
coupled in the infrared.  This is a well-motivated assumption if the vectorial
gauge theory arises as a low-energy effective field theory from an ultraviolet
completion which is a chiral gauge theory.  This is natural if the latter
theory becomes strongly coupled, since it can then form fermion condensates
that self-break it down to the vectorial subgroup symmetry.  However, one may
also choose to focus on the vectorial gauge theory as an ultraviolet-complete
theory in itself. In a vectorial gauge theory, an intrinsic (bare) mass term
for a fermion $\psi$, ${\cal L}_m = - m\bar\psi\psi$, is allowed by the gauge
invariance.  (For an SU(2) theory, with fermions written as left-handed chiral
fields, the gauge-invariant mass term can be expressed in a Majorana form,
e.g., for the fundamental representation, $m' \epsilon_{ij} \psi^{i \ T}_L C
\psi^j_L$.)  Hence, one may consider a more general situation in which the
fermions may have such intrinsic masses in the high-scale Lagrangian.  Quantum
chromodynamics (QCD) provides an example of this, in which the quarks have hard
(also called current-quark) masses \cite{hard} that span a large range, from
$m_u$ of a few MeV to $m_t \simeq 172$ GeV. In particular, this range extends
both far below and far above, the scale $\Lambda_{QCD} \simeq 300$ MeV where
the QCD coupling $\alpha_s(\mu)$ becomes O(1) and the theory confines and
spontaneously breaks chiral symmetry.

The main effect of intrinsic fermion masses here is the same as in QCD; as the
reference scale $\mu$ decreases below the value of such a mass of some fermion
$m_f$, the beta function changes from one that includes this to one that
excludes this in the set of light, active fermions.  For a theory with a set
$\{N_R\}$ such that $b_2 < 0$ at a high scale, and hence evolution toward an
approximate or exact infrared fixed point, the reduction of one or more numbers
$N_R$ can reverse the sign of $b_2$, making it positive and removing this
infrared fixed point.  Indeed, in principle, a theory could have sufficiently
large numbers of fermions in various representations $\{N_R\}$ that it is not
asymptotically free at a high energy scale above the fermion masses, but as
this scale decreases below some of these masses, the modified beta function
describing the gauge coupling evolution in the result low-energy effective
field theory is asymptotically free.

\section{Conclusions}

In this paper we have studied the evolution of an asymptotically free vectorial
SU($N$) gauge theory from high scales to the infrared and the resultant phase
structure in the general case in which the theory contains fermions
transforming according to several different representations of the gauge group.
Using information from the beta function and results from an approximate
analysis of the Dyson-Schwinger equation for the fermion(s), we have
investigated examples that illustrate a wide range of possible behavior.  In
one type of model, the theory contains sufficiently few fermions that the
coupling $\alpha$ increases as the reference scale decreases, but the 2-loop
beta function does not have an infrared zero away from the origin.  In this
case, as $\alpha$ increases and exceeds a critical value for the formation of a
condensate of fermions with the largest $C_2(R)$, this forms, the fermions gain
dynamical masses, and these fermions are then integrated out of the low-energy
effective field theory applicable below this highest condensation scale.  In
the low-energy theory, the coupling $\alpha$ continues to evolve, but according
to a different beta function, and there is then condensation of the fermions
with the next largest value of $C_2(R)$, and so forth. In another type of
model, the theory contains enough fermions in various representations that the
beta function does have an infrared zero.  In this case, there are two main
categories of behavior.  In one type, the value of $\alpha_{IR}$ is larger than
the critical value for condensation of the fermions with the largest $C_2(R)$,
so this condensation occurs, and is followed by sequential condensation(s) at
lower scales.  In a second type, the value of $\alpha_{IR}$ is sufficiently
small that there are no condensates formed, there is no spontaneous chiral
symmetry breaking, and $\alpha_{IR}$ is an exact infrared fixed point of the
renormalization group.  We have given explicit examples of each of these types
of behavior in the case of an SU(2) gauge theory. We have also briefly
discussed the effects of nonzero intrinsic fermion masses. 

We thank T. Appelquist and F. Sannino for helpful discussions.  This research
was partially supported by the grant NSF-PHY-06-53342.

\begin{table}
\caption{\footnotesize{Some numerical results for the SU(2) theory. IRFP
    denotes an (exact or approximate) infrared fixed point of the
    renormalization group equation for $\alpha$. nIRFP means that the two-loop
    beta function does not have such an IRFP, i.e., a zero away from the
    origin.  In the columns marked $c_I$ for $I=1/2, \ 1$ we indicate with a y
    (yes) or n (no) whether the $\beta DS$ method with the one-gluon (ladder)
    approximation to the DS equation, predicts that there is condensation of
    the isospin $I$ fermions. The notation $m$ means ``maybe'', reflecting the
    substantial theoretical uncertainties in the $\beta DS$ predictions due to
    the strong-coupling nature of the physics. If the theory with all of its
    massless fermions has a IRFP, this is marked as $\alpha_{IR,h}$, where $h$
    stands for ``highest-scale''.  If the low-energy effective field theory
    applicable for energies below the highest condensation scale has an IRFP,
    this is denoted $\alpha_{RI,\ell}$, where $\ell$ stands for ``lower
    scale''~. In cases where no condensation occurs for any of the isospin $I$
    fermions, $\alpha_{IR,h}=\alpha_{IR,\ell}$.}}
\begin{center}
\begin{tabular}{|c|c|c|c|c|c|} \hline\hline
$N_{1/2}$ & $N_{1}$ & $\alpha_{IR,h}$ & $\alpha_{IR,\ell}$ & $c_{1/2}$ & $c_1$
\\ \hline
    1     &    0    &   nIRFP        &    $-$         &   y       &  $-$  \\
    2     &    0    &   nIRFP        &    $-$         &   y       &  $-$  \\
    3     &    0    &   nIRFP        &    $-$         &   y       &  $-$  \\
    4     &    0    &   nIRFP        &    $-$         &   y       &  $-$  \\
    5     &    0    &   mIRFP        &    $-$         &   y       &  $-$  \\
    6     &    0    &   11.4         &    $-$         &   y       &  $-$  \\
    7     &    0    &   2.83         &    $-$         &   y       &  $-$  \\
    8     &    0    &   1.26         &    $-$         &   m       &  $-$  \\
    9     &    0    &   0.59         &    $-$         &   n       &  $-$  \\
   10     &    0    &   0.23         &    $-$         &   n       &  $-$  \\ 
\hline    
    0     &    1    &   nIRFP        &    $-$         & $-$        & y    \\
    1     &    1    &   9.14         &   IRFP         & y          & y    \\
    2     &    1    &   3.06         &   IRFP         & y          & y    \\
    3     &    1    &   1.53         &   nIRFP        & y          & y    \\
    4     &    1    &   0.84         &   nIRFP        & y          & y    \\
    5     &    1    &   0.44         &   mIRFP        & m          & m    \\
    6     &    1    &   0.18         &   0.18         & n          & n    \\
\hline
    0     &    2    &   1.26          &    $-$         &  $-$      &  y   \\
    1     &    2    &   0.59          &   nIRFP        &   y       &  m   \\
    2     &    2    &   0.23          &   0.23         &   n       &  n   \\
\hline\hline
\end{tabular}
\end{center}
\label{su2table}
\end{table}


\begin{thebibliography}{99}

\bibitem{rn}
%
No confusion should result from the use of $R$ for both representation and the
right-handed chiral component of a fermion; the meaning will be clear from the
context.

\bibitem{higherrep}
%
K. Lane and E. Eichten, Phys. Lett. B {\bf 222}, 274 (1989); D. Hong, S. Hsu,
and F. Sannino, Phys. Lett. B {\bf 597}, 89 (2004); F. Sannino and K. Tuominen,
Phys. Rev. D {\bf 71}, 051901(R) (2005); N. D. Christensen and R. Shrock,
Phys. Lett. B {\bf 632}, 92 (2006); D. D. Dietrich and F. Sannino, Phys. Rev. D
{\bf 75}, 085018 (2007); R. Foadi, M. T. Frandsen, T. A. Ryttov,
and F. Sannino, Phys. Rev. D {\bf 76}, 055005 (2007); F. Sannino, Phys. Rev. D
{\bf 79}, 096007 (2009); T. A. Ryttov and F. Sannino, arXiv:0906.0307; 
H. S. Fukano and F. Sannino, ArXiv:1005.3340;
T. A. Ryttov and R. Shrock, ArXiv:1005.3844.

\bibitem{sanrev}
F. Sannino, ArXiv:0911.0931.

\bibitem{colorrep}
See, e.g., G. Karl, Phys. Rev. D {\bf 14}, 2374 (1976); 
W. J. Marciano, Phys. Rev. D {\bf 21}, 2425 (1980) and references therein. 

\bibitem{b1}
D. J. Gross and F. Wilczek, Phys. Rev. Lett. {\bf 30}, 1343 (1973); 
H. D. Politzer, Phys. Rev. Lett. {\bf 30}, 1346 (1973); G. 't Hooft,
unpublished. 

\bibitem{b2}
W. E. Caswell, Phys. Rev. Lett. {\bf 33}, 244 (1974); 
D. R. T. Jones, Nucl. Phys. B {\bf 75}, 531 (1974). 

\bibitem{gp}
%
Our normalizations for the quadratic Casimir and trace invariants of a Lie
group are standard.  The quadratic Casimir invariant $C_2(R)$ for the
representation $R$ is given by $\sum_{a=1}^{o(G)} \sum_{j=1}^{dim(R)}
[D_R(T_a)]_{ij} [D_R(T_a)]_{jk} = C_2(R)\delta_{ik}$, where $a,b$ are group
indices, $o(G)$ is the order of the group, $T_a$ are the generators of the
associated Lie algebra, and $D_R(T_a)$ is the matrix form of the $T_a$ in the
representation $R$.  The trace invariant $T(R)$ is defined by 
$\sum_{i,j=1}^{dim(R)}[D_R(T_a)]_{ij} [D_R(T_b)]_{ji} = T(R)\delta_{ab}$.

\bibitem{bz}
T. Banks and A. Zaks, Nucl. Phys. B {\bf 196}, 189 (1982). 

\bibitem{lmax}
S. Brodsky and R. Shrock, Phys. Lett. B {\bf 666}, 95 (2008).

\bibitem{btd}
S. J. Brodsky, G. F. de T\'eramond, and A. Deur, ArXiv:1002.3948. 

\bibitem{creutz}
M. Creutz, unpublished. 

\bibitem{4f}
%
See, e.g., M. Kurachi, R. Shrock, and K. Yamawaki, 
Phys. Rev. D {\bf 76}, 035003 (2007); F. Sannino, ArXiv:1003.0289, and
references therein. 

\bibitem{lane}
K. Lane, Phys. Rev. D {\bf 10}, 1353, 2605 (1974).

\bibitem{politzer}
H. D. Politzer, Nucl. Phys. B {\bf 117}, 397 (1976).

\bibitem{wtc}
T. Appelquist, D. Karabali, and L. C. R. Wijewardhana, Phys. Rev. Lett. {\bf
57}, 957 (1986); T. Appelquist and L. C. R. Wijewardhana, Phys. Rev. D
{\bf 35}, 774 (1987);  Phys.  Rev. D {\bf 36}, 568 (1987).

\bibitem{bds}
T. Appelquist, J. Terning, and L. C. R. Wijewardhana,
Phys. Rev. Lett. {\bf 77}, 1214 (1996). 

\bibitem{alm}
T. Appelquist, K. Lane, and U. Mahanta, Phys. Rev. Lett. {\bf 61}, 1553 
(1988).

\bibitem{otherw}
B. Holdom, Phys. Lett. {\bf B150}, 301 (1985); K Yamawaki,
M. Bando, and K.  Matumoto, Phys. Rev. Lett. {\bf 56}, 1335 (1986). 

\bibitem{ksssu2}
J. B. Kogut, M. Stone, H. W. Wyld, S. H. Shenker, J. Shigemitsu, and D. K.
Sinclair, Nucl. Phys. B225, 326 (1983); J. B. Kogut, J. Shigemitsu, and 
D. K. Sinclair, Phys. Lett. B138, 283 (1984).

\bibitem{ksssu3}
J. B. Kogut, M. Stone, H. W. Wyld, J. Shigemitsu, S. H. Shenker, D. K.
Sinclair, Phys. Rev. Lett. 48. 1140 (1982); J. B. Kogut, J. Shigemitsu, and 
D. K. Sinclair, Phys.Lett. B145. 239 (1984).

\bibitem{iwasaki}
Y. Iwasaki, K. Kanaya, S. Sakai, and T. Yoshié, Phys. Rev. Lett. {\bf 69}, 21
(1992); Y. Iwasaki et al., Phys. Rev. D {\bf 69}, 014507 (2004).

\bibitem{heller}
P. H. Damgaard, U. M. Heller, A. Krasnitz, and P. Olesen, Phys. Lett. B 
{\bf 400}, 169 (1997). 

\bibitem{afn}
%
T. Appelquist, G. Fleming, and E. Neil, Phys. Rev. Lett. {\bf 100}, 171607
(2008); Phys. Rev. D {\bf 79}, 076010 (2009); T. Appelquist et al.,
Phys. Rev. Lett. {\bf 104}, 071601 (2010); 
A. Deuzeman, M. P. Lombardo, and E. Pallante, Phys. Lett. B {\bf 670}, 41
(2008); arXiv:0810.1719, arXiv:0904.4662; 
X.-Y. Jin and R. Mawhinney, PoS Lattice-2008:059 (2008), arXiv:0812.0413;
A. Hasenfratz, ArXiv:1004.1004. 

\bibitem{lgtsym}
%
S. Catterall, J. Giedt, F. Sannino, and J. Schneible, JHEP 0811, 009 (2008);
ArXiv:0910.4387; 
T. DeGrand, Y. Shamir, and B. Svetitsky, Phys. Rev. D {\bf 79}, 
Phys. Rev. D {\bf 79}, 034501 (2009);
Z. Fodor, K. Holland, J. Kuti, D. Nogradi, and C. Schroeder,
Phys. Lett. B {\bf 681}, 353 (2009); JHEP {\bf 2009}, 103 (2009); 
A. J. Hietanen, K. Rummukainen, and K. Tuominen, 
Phys. Rev. D {\bf 80}, 094504 (2009); 
L. Del Debbio, B. Lucini, A. Patella, C. Pica, and A. Rago,
Phys. Rev. D {\bf 80}, 074507 (2009); 
F. Bursa, L. Del Debbio, L. Keegan, C. Pica, and T. Pickup,
Phys. Rev. D {\bf 81}, 014505 (2010). 
J. B. Kogut and D. K. Sinclair, ArXiv:1002.2988. 

\bibitem{tc}
S. Weinberg, Phys. Rev. D {\bf 19}, 1277 (1979); 
L. Susskind, Phys. Rev. D {\bf 20}, 2619 (1979).

\bibitem{etc}
S. Dimopoulos and L. Susskind, Nucl. Phys. {\bf B155}, 237 (1979); 
E. Eichten and K. Lane, Phys.  Lett. {\bf B90}, 125 (1980).

\bibitem{at94}
Some examles are 
T. Appelquist and J. Terning, Phys. Rev. D {\bf 50}, 2116 (1994); 
T. Appelquist and R. Shrock, Phys. Lett. B {\bf 548}, 204 (2002);
T. Appelquist and R. Shrock, Phys. Rev. Lett. {\bf 90}, 201801 (2003);
T. Appelquist, M. Piai, and R. Shrock, Phys. Rev. D {\bf 69}, 015002 (2004);
T. A. Ryttov and R. Shrock, ArXiv:1004.2075. 

\bibitem{topcolor}
C. Hill and E. Simmons, Phys. Rep. {\bf 381}, 235 (2003);
A. Martin and K. Lane, Phys. Rev. D {\bf 71}, 015011 (2005).

\bibitem{pdg}
R. S. Chivukula, M. Narain, and J. Womersley, Phys. Lett. B {\bf 667},
1258 (2008) and http://pdg.lbl.gov. 

\bibitem{dewsb}
See talks in {\it Workshop on Dynamical Electroweak Symmetry Breaking},
Southern Denmark University, 2008, http://hep.sdu.dk/dewsb. 

\bibitem{thooft79}
G. `t Hooft, in {\it Recent Developments in Gauge Theories} 
(1979 Carg\` ese Summer Institute) (Plenum, New York, 1980), p. 135.

\bibitem{casher}
A. Casher, Phys. Lett. B {\bf 83}, 395 (1979). 

\bibitem{seiberg}
%
Reviews include K. Intrilligator and N. Seiberg, in D. E. Soper, ed., {\it QCD
and Beyond: Proceedings of the Theoretical Advanced Study Institute in
Elementary Particle Physics, 1995} (World Scientific, Singapore, 1996), p. 223
(hep-th/9509066); M. Shifman, Prog. Part. Nucl. Phys. {\bf 39}, 1 (1997)
(hep-th/9704114).

\bibitem{acs}
T. Appelquist, A. G. Cohen, and M. Schmaltz, 
Phys. Rev. D {\bf 60}, 045003 (1999).  

\bibitem{acss}
T. Appelquist, A. G. Cohen, M. Schmaltz, and R. Shrock, 
Phys. Lett. B {\bf 459}, 235 (1999). 

\bibitem{thooft}
G. 't Hooft, Phys. Rev. Lett. {\bf 37}, 8 (1976); Phys. Rev. D {\bf 14},
3432 (1978).

\bibitem{c2n}
%
Note that, in contrast to the case of SU(2), where the value of $C_2(I)$
increases monotonically with the dimension ${\rm dim}(R_I) = 2I+1$, this is not
the case with SU(3); for example, the rank-2 symmetric representation with
$(p_1,p_2)=(2,0)$ has a smaller dimension, ${\rm dim}(R_{(2,0)} \equiv
d(2,0)=6$, than the adjoint representation (1,1), with $d(1,1)=8$, but has a
larger value of $C_2$: $C_2(2,0) = 10/3$, while $C_2(1,1)=3$.

\bibitem{hard}
Here a ``hard'' quark mass is a mass that would remain in the hypothetical
limit in which one turned off color interactions.  If these masses are
dynamically generated, then they are actually soft on the higher scale where
they arise \cite{sml}. 

\bibitem{sml}
N. D. Christensen and R. Shrock, Phys. Rev. Lett. {\bf 94}, 241801 (2005). 

\bibitem{nt}
T. Appelquist and R. Shrock, Phys. Lett. B {\bf 548}, 204 (2002);
Phys. Rev. Lett. {\bf 90}, 201801 (2003).

\bibitem{ckm}
T. Appelquist, M. Piai, and R. Shrock, Phys. Rev. D {\bf 69}, 015002 (2004).

\end{thebibliography}
\end{document}